\documentclass[structabstract]{aa}

\usepackage{graphicx}
\usepackage{txfonts}

\begin{document}


\title{GRB Afterglows with Energy Injection from a spinning down NS}


\author{S. Dall'Osso\inst{1,2} \and G. Stratta\inst{3}
\and D. Guetta\inst{1} \and S. Covino\inst{4} 
\and G. De Cesare\inst{5,6,7}
\and L. Stella\inst{1}
}


\institute{INAF - Osservatorio Astronomico di Roma, via Frascati 33,
Monte Porzio Catone, Roma, Italy 
\and Virgo-Ego Scientific Forum fellow
\and ASDC, via Galileo Galilei, 00040, Frascati (Roma), Italy\thanks{INAF
personnel resident at ASDC} 
\and  INAF - Osservatorio Astronomico di Brera, Via E. Bianchi 46, 23807, 
Merate (LC), Italy
\and INAF - Istituto di Astrofisica Spaziale e Fisica Cosmica di Roma, via
Fosso del Cavaliere 100, I-00133 Roma, Italy
\and Dipartimento di Astronomia, Universita' degli Studi di Bologna, Via
Ranzani 1, I40127 Bologna, Italy
\and Centre d'Etude Spatiale des Rayonnements, CNRS/UPS, B.P. 4346, 31028
Toulouse Cedex 4, France
}


\abstract
{}
{
We investigate a model for the shallow decay phases 
of Gamma-ray Burst (GRB) afterglows discovered by Swift/XRT in the first hours 
following a GRB event. In the context of the fireball scenario, 
we consider the possibility that long-lived energy injection from 
a millisecond spinning, ultramagnetic neutron star (magnetar) powers afterglow 
emission during this phase.}
{We consider the energy evolution in a relativistic shock subject to both 
radiative losses and energy injection from a spinning down magnetar in 
spherical symmetry. We model the energy injection term through magnetic dipole 
losses and discuss an approximate treatment for the dynamical evolution of the 
blastwave. 
We obtain an analytic solution for the energy evolution in the 
shock and associated lightcurves. To fully illustrate the potential 
of our solution we calculate lightcurves for a few selected X-ray afterglows 
observed by Swift and fit them using our theoretical lightcurves.} 
{Our solution naturally describes in a single picture the properties of the 
shallow decay phase and the transition to the so-called normal decay phase. In 
particular, we obtain remarkably good fits to X-ray afterglows for plausible
parameters of the magnetar. Even though approximate, our 
treatment provides a step forward with respect to previously adopted 
approximations and provides additional support to the idea that a millisecond 
spinning (1-3 ms), ultramagnetic (B$\sim 10^{14}-10^{15}$ G) neutron star 
loosing spin energy through magnetic dipole radiation can explain the 
luminosity, durations and shapes of X-ray GRB afterglows.} 
{}


\keywords{(Stars:) Gamma-ray burst: general --- X-rays: bursts --- Shock Waves 
--- Stars: magnetars --- Relativistic processes}

\maketitle

\section{Introduction}
\label{introduction}

Before the launch of Swift in November 2004, X-ray afterglows of long
Gamma-Ray Bursts could be pointed with X-ray telescopes 
not earlier than several hours after the 
trigger. These observations showed in most cases a smooth power-law like decay 
$F(t)\propto t^{-\alpha}$,
with typical index of $\alpha\geq 1$. 
With the advent of Swift, X-ray fluxes could be monitored from a few minutes  
after the burst trigger. These observations have revealed a complex behavior 
in the first few hours after the GRB, which nonetheless displays remarkably standard 
properties across different events. This behavior can be described 
with a double broken power law, with an initial very steep decay (up to few 
hundreds of seconds after the trigger) with $\alpha>2$ followed by a 
shallow phase, lasting $\sim 10^3-10^4$ s, with $\alpha<0.8$ and, later on, a 
steeper 'normal' decay with $\alpha\sim1.2-1.4$ (Nousek et al. 2006, 
Gehrels et al. 2009).  

The X-ray spectral slope does not change between the shallow and normal decay, 
in marked contrast to what would be expected in case this temporal break was 
caused by
the passage of a characteristic synchrotron frequency in the X-ray band 
(e.g. Sari et al. 1998). A possible interpretation requiring no spectral 
variations in the observed energy band invokes prolonged energy injection into 
the external shock that is believed to give rise to the GRB afterglow. Energy 
injection could come either from relativistic shells impacting the 
fireball at late times (e.g. Rees and Meszaros 1998, Sari and Meszaros 2000) 
or from a long-lived central engine (e.g. Zhang et al. 2006, Nousek et al. 
2006, Panaitescu et al. 2006a).

Among different hypotheses on the nature of GRB central engines, two major 
classes can be identified. The first considers the formation of a black hole-
debris torus system, the prompt emission being related to accretion of 
matter from the torus during the first $\sim 10-100$s (Narayan, Paczynski
\& Piran 1992, Woosley 1993, Meszaros, Rees \& Wijers 1999). In this scenario,
keeping the energy production active 
in order to power the afterglow for at 
least $\sim 10^4$ s is a difficult and far from 
settled matter, 
(Mc Fadyen et al. 2001, Ramirez-Ruiz 2004, Cannizzo 
\& Gehrels 2009, Barkov \& Komissarov 2009).

An alternative class of models invokes the formation of a strongly 
magnetic (B$>10^{14}-10^{15}$ G) , millisecond spinning neutron star (NS, 
Usov 1992, Duncan \& Thompson 1992, Blackman \& Yi 1998, Kluzniak \& Ruderman 1998,
Wheeler 2000). Recently, time-dependent MHD simulations have shown that long 
GRBs can originate from the interaction between a relativistic and 
strongly magnetized wind produced by a newly-born NS and the surrounding 
stellar envelope.
NS spin periods of $\sim 1$ ms and ultrastrong magnetic 
fields, \textit{i.e.} B$\geq 10^{15}$ G, would be required 
in this case  (e.g. Thompson et al. 2004, 
Bucciantini et al. 2006, 2008, 2009; see also Tchekhovskoy, McKinney \& 
Narayan 2009). 

The newly formed NS is expected to 
loose its initial spin energy ($>10^{52}$ erg) at a very high rate for the first
few hours through magnetic dipole spin down, something that provides a 
long-lived central engine in a very natural way. 
Dai \& Lu (1998) considered this idea in relation to possible observable 
effects on the afterglow emission. Zhang \& Meszaros (2001) 
argued that, in this scenario, achromatic bumps in afterglow lightcurves are 
expected for NS spin periods shorter than a few ms and magnetic fields
stronger than several times $10^{14}$ G. 
Interestingly, studies of the origin of NS magnetism envisage that millisecond 
spin period at birth is the key property that allows a proto-NS to amplify a 
seed magnetic field to a strength far exceeding $10^{14}$ G, through efficient 
conversion of its initial differential rotation energy (e.g Duncan \& Thompson 
1992, Thompson \& Duncan 1993). Such highly magnetized, fast spinning NSs are
expected to loose angular momentum at a high rate 
in the first decades of their life and later 
become slowly rotating magnetars whose major free energy reservoir is 
in their magnetic field (Thompson \& Duncan 1995, 1996, 2001, cfr. Woods \& 
Thompson 2006, Mereghetti 2008). We term these NSs as magnetars since their 
birth even though, when they spin at millisecond period, their rotational 
energy is still the main free energy reservoir.

After the Swift discovery of early afterglow shallow phases, the magnetar 
scenario has been invoked to interpret the X-ray light curve of both 
some short and 
long GRBs (e.g. 051221A by Fan and Xu 2006; 060313 by Yu and Huang 2007; 
GRB 050801 by De Pasquale et al. 2007; 070110 by Troja et al. 2007).  For GRB 
060729 this scenario was shown to provide a good agreement with the shallow 
and normal decay phases in the optical and X-ray bands (Grupe et al.
2007, Xu et al. 2009).\\
%
Finally we note that, besides the interest in understanding GRB physics, the 
very fast spin and huge magnetic field envisaged in the magnetar formation
scenario makes these objects very interesting also for gravitational wave (GW)
astronomy. Different possibilities for this to occur have been investigated
in the literature (Palomba 2001, Cutler 2002, Stella et al. 2005, 
Dall'Osso \& Stella 2007, Dall'Osso, Shore \& Stella 2009, Corsi \& Meszaros
2009) showing that, in astrophysically plausible conditions, GW emission might
efficiently extract spin energy from the NS, in competition with magnetic 
dipole losses. 
The study presented in this paper builds on the ansatz
that millisecond spinning
magnetars are formed in the events that give rise to long GRBs. 
We investigate the evolution of energy in a relativistic blastwave 
subject to radiation losses due to shock deceleration in the ISM and energy 
injection from a magnetically braking NS. We extend previous treatments by
describing the injection term by the standard magnetic dipole formula and 
deriving a prediction for the evolution of energy and luminosity that can 
interpret the X-ray afterglows through their shallow and normal decay phases 
altogether. We derive an approximate solution for the blastwave 
luminosity 
which we compare with X-ray GRB afterglow lightcurves observed by Swift. 
We obtain a remarkably good match to these lightcurves
for the range of initial spin periods and magnetic field
strengths expected for magnetars at birth. These results illustrate the
potential of this scenario in explaining the early afterglow observations in a
simple, unified picture. 
%
\section{Relativistic Blast Wave with Energy Injection: spherically symmetric
case}
\label{model}


We assume that a GRB event is associated to the formation of a millisecond
spinning, ultramagnetized NS.
In the context of the fireball scenario, the energy released in the collapse of 
the progenitor star produces first a fireball 
expanding freely at relativistic speed through the ambient medium. 
The prompt emission 
is produced at this early stage and is commonly ascribed to internal shocks 
in the fireball (Rees \& Meszaros 1994,  Paczynski \& Xu 1994, Sari \& 
Piran 1997). A relativistic forward shock is produced at larger distances
from the explosion site ($\sim 10^{16}$ cm), which initially propagates 
freely through the ambient medium.
At a later time, call it $t_d$, the mass swept up by the forward 
shock will be enough to begin affecting the expansion dynamics of the shock 
itself. This defines the decelaration radius $r_d \approx ct_d$, at which the 
kinetic energy of the shock starts being efficiently converted to internal 
energy and then radiation. This corresponds to the onset of the afterglow 
emission. 
We focus here only on the deceleration phase, describing the evolution of the 
total energy within the fireball as matter from the ISM is swept up. Our aim is 
to interpret the shallow decay phase
and subsequent achromatic transition to the ``normal'' decay phase as
observed in X-rays, within a single physical model containing a minimal set of 
parameters. We do not address here a detailed study of the 
multiwavelength behaviour of afterglow lightcurves. 
In $\S$ \ref{further} we discuss possible developments of our work in this
direction, as to closely compare model predictions with multiwavelength 
observations.
The first few minutes after the GRB event are characterized by a very steep 
power-law decay of the flux while a marked spectral change usually accompanies 
the transition to the shallow decay phase (this is in contrast with the lack 
of spectral evolution across the shallow-to-normal transition).
This initial steep decay is believed to arise from a different spectral 
component than the X-ray afterglow, likely the tail of the prompt emission 
(cfr. Zhang 2007 for a detailed discussion);  
we do not consider it in this work.

In addition to deceleration in the ISM, we study the way in which the 
afterglow emission is affected by the energy injection caused by the spindown 
of the newly formed magnetar. 
We first introduce time $t$ as that measured by a clock at rest in 
the NS (central engine) frame. In this frame the NS loses rotational energy, 
likely in the form of a strongly magnetized particle wind, with a luminosity 
L$_{sd}$(t) according to the usual magnetic dipole spindown formula 
\begin{equation}
\label{Lsd}
L_{sd} (t) = \frac{I~K \omega^4_i}{(1+2 K\omega^2_i t)^2} = 
\frac{L_i}{(1+at)^2} = \frac{E_{s,i}}
{t_2 (1+t/t_2)^2}~, 
\end{equation}
%
where $I$ is the NS moment of inertia, $K$ = $B^2R^6$/(6$Ic^3$) with $B$ the 
(dipole) magnetic field at the NS pole, $R$ the NS radius and $c$ the speed of 
light. In the second equality, the quantity $L_i =$ $L_{sd}$($t_i$) 
represents the spindown luminosity at the initial time (t$_i$) when spindown 
through magnetic dipole radiation sets in, and $a$ = 2 $K
\omega^2_i = 1/t_2$, where $t_2$ represents the spindown timescale at time 
$t_i$ and $\omega_i$ is the initial spin frequency. $E_{s,i}$ is the NS spin 
energy at time $t_i$, so that $L_i = E_{s,i}/t_2$. The energy carried by the 
wind travels essentially at the speed of light,
so that the energy emitted at later times by the 
NS can be 
transferred to the shock. 

To calculate the expected behavior of the lightcurve 
we start from the energy balance of the relativistic blastwave subject to 
the energy injection in eq. (\ref{Lsd}) along with radiative losses. The latter 
are described by following the prescription 
of Cohen, Sari \& Piran (1998). For the time being we assume spherical 
symmetry of all processes involved, which allows us to write
%
the complete energy equation of the blast wave as
%
\begin{equation}
\label{energyone}
\frac{dE}{dt} = L_{inj}(t) - k\frac{E}{t} = (1-\beta) 
L_{sd}[t-\frac{r(t)}{c}] - k \frac{E}{t}~.
\end{equation}
Here $k= 4 \epsilon_e$, with $\epsilon_e$ the fraction of the total energy that 
is transferred to the electrons, $r(t)$ is the radius of the blast wave at time 
$t$ and all quantities are expressed in the frame of the central engine. 
Note that L$_{inj}$ represents the rate at which energy is injected in 
the shock at time $t$. This quantity is related to the rate at which the 
central NS emitted energy - L$_{sd}$ - at a previous time, $t-r(t)/c$.
In the central engine rest frame, an infinitesimal time interval 
$dt$ is related to the infinitesimal displacement of the blast wave $dr = 
c \beta(t) dt$. 
However, due to propagation effects, photons emitted at two successive radii 
will be received by the observer over a much shorter time interval d$T$, which 
defines what is called ``the observer's time'' ($T$).
The relation between $dt$ and $dT$ is\footnote{the second equality holds 
since $\beta \simeq 1$}
\begin{equation}
\label{transformtime1}
dT = (1-\beta) dt \simeq \frac{dt}{2\Gamma^2(t)}~.
\end{equation}
When integrated, this gives $T = t -r(t)/c$.
%
%
Now we can transform eq. (\ref{energyone}) to the equivalent form with respect 
to time $T$, using relation (\ref{transformtime1}). 
After some manipulation one obtains
\begin{equation}
\label{energytwo}
\frac{dE}{dT} = 
L_{sd}(T) - k \frac{E}{T} \left(\frac{\mbox{\small{dln}}t}
{\mbox{\small{dln}}T}\right)~.
\end{equation}
%
%
%
%
In order to obtain $t(T)$, $\Gamma(t)$ is required (see eq. 
\ref{transformtime1}); this in turn requires a study
of the hydrodynamical evolution of the blast wave. 
The solution to this problem with energy injection is far from trivial and 
beyond the scope of the present paper. Here we introduce an approximation in 
order to derive a solution to the problem which captures all the essential 
physics. Customarily the evolution of $\Gamma$ in the deceleration phase is 
treated using self-similar solutions for relativistic blastwaves (Blandford 
\& McKee 1976, Piran 1999). In this case all quantities scale as power-laws 
(with time $t$). 
Upon writing $\Gamma^2\propto t^{-m}$ one can solve eq. (\ref{transformtime1}) 
and obtain $T = 
t/[2(m+1)\Gamma^2]$, from which dln$t/$dln$T = 1/(1+m)$ can be substituted in 
eq. (\ref{energytwo}) to obtain $E(T)$. On the other hand, in the problem we 
are considering the shock evolution is not self-similar since the energy 
injection term introduces in the problem the timescale $t_2 = a^{-1}$. 
One can evaluate the change introduced by this complication using an
integral expression for the total (internal plus kinetic) energy of the shock
(Zhang \& Meszaros 2001)
\begin{equation}
\label{closure_energy}
E = \frac{4\pi}{3} n (m_p c^2) \Gamma^2 r^3 \approx  \frac{4\pi}{3} n (m_p c^5) 
\Gamma^2 t^3~,
\end{equation}
where $r \approx ct$ has been assumed. 
By neglecting, for the sake of simplicity, radial variations in the density of the
ambient medium, 
we can thus write $\Gamma$ as a function of the shock energy and time $t$, 
namely $\Gamma^2 \propto E/t^3$. For self-similar solutions one obtains 
%
E$\propto t^{3-m}$; adiabatic shocks thus
correspond to $m=3$.
The relation between
$\Gamma$ and $E$ expressed by eq. (\ref{closure_energy}) identifies 
two extremes for the evolution of $\Gamma$.
First, neglecting radiative 
losses correponds to the fastest possible growth rate for $E$ which, in 
turn, corresponds to the slowest decay rate for $\Gamma^2$. On the other hand, 
neglecting energy injection corresponds to the fastest possible decay rate 
for $E$ and, in turn, of $\Gamma^2$. Any realistic behaviour of $\Gamma$ is 
thus expected to lie between these two extremes. 

In the former case, which is appropriate for
the early stages of energy injection, we can write to a very good approximation
\begin{equation}
\label{extreme1} 
\frac{dE}{dt} \approx \frac{\mbox{\small{const}}}{\Gamma^2}~.
\end{equation}
%
The solution to this
gives E$\propto t^2$ and, thus, $\Gamma^2 \propto t^{-1}$ or dln$t$/dln$T$ = 
1/2. In the opposite extreme, where only radiation losses are present, one 
obtains E$\propto t^{-k}$ which implies $\Gamma^2 \propto t^{-(3+k)}$ (with, 
in general, $k <1$) or dln$t$/dln$T = 1/(4+k)$. Note that these
extremes reproduce, as expected, the self-similar solution obtained by Zhang 
\& Meszaros 2001.\\ 
Hence, although the coefficient multiplying $E/T$ in eq. (\ref{energytwo}) does 
depend on time, its value will be bracketed between the two extremes found 
above\footnote{Note that, for a wind-like medium whose density declines as 
$r^{-2}$, similar conclusions would hold: adiabatic shocks correspond to $m=1$, 
while dln$t$/dln$T$ = 1 and $1/(1+k)$ for the two extremes, respectively.}, as 
long as $E \propto \Gamma^2 t^3$ holds. As these extremes differ by a factor 
$2+k/2$ at most, we can consider dln$t$/dln$T \approx$ const as a reasonable 
first-order approximation, neglecting the slow and moderate change of 
dln$t$/dln$T$. 
This allows us to write the energy equation as
%
\begin{equation}
\label{equationabove}
\frac{dE}{dT} = L_{sd}(T) - k' \frac{E}{T}~,
\end{equation}
where $k' = k($dln$t$/dln$T$)$\approx$ const. Note that its value will also 
depend on the unknown density profile of the ambient medium, 
about which we do not make any assumptions. 
\begin{figure}
\includegraphics[scale=0.39, angle=-90]{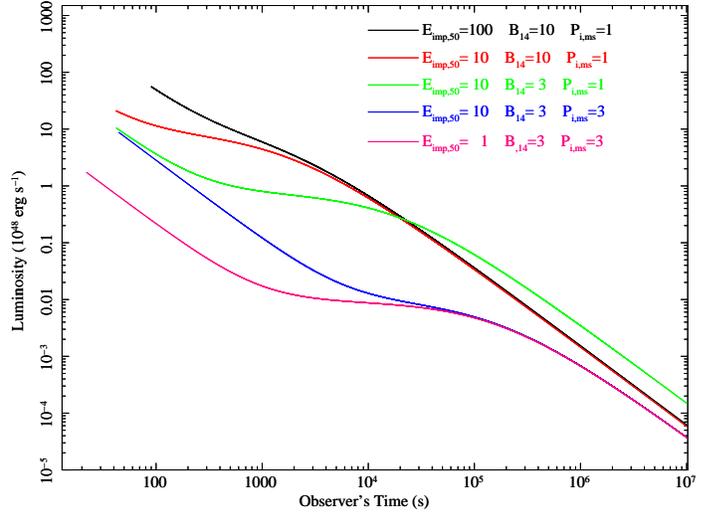}
\caption{Five different theoretical (bolometric) lightcurves from the model
presented here, drawn varying the initial energy of the afterglow 
(E$_{\mbox{\tiny{imp}}}$), the dipole magnetic field (B) and the initial spin 
period (P$_{\mbox{\tiny{i,ms}}}$) of the NS. All lightcurves are obtained for the 
same value of $k'=0.4$. The time at which each lightcurve begins is the 
deceleration time  $T_d$, estimated for illustration purposes simply by 
equating the initial energy E$_{\mbox{\tiny{imp}}}$ to the rest-mass 
energy of swept up matter in a constant density ISM ($n \simeq 1$ cm$^{-3}$).}
\label{fig1}
\end{figure}
Our ignorance about it is completely contained in the free parameter $k'$, as 
is our ignorance about the microphysics. In general, fixing all other 
parameters, we expect larger values of $k'$ for a wind-like medium than for a 
constant density ISM, based on the discussion above. 
As far as our present work is concerned, the solution to eq. 
(\ref{equationabove}) can be cast in the form
%
\begin{equation}
\label{formally-solved}
E(T) = \frac{L_i}{T^{k'}} \int_{T_0}^T \frac{T^{k'}}{(1+aT)^2}
~dT + E_0\left(\frac{T_0}{T}\right)^{k'}~,
\end{equation}
where T$_0 \geq$ T$_d$ is any time chosen as the initial condition. 
%
%
The integral in the above expression 
can be expressed in terms of the (real valued) hypergeometric function $_2F_1
(a, b; c; z)$ 
%
\begin{equation}
\label{hypergeom}
\int \frac{T^{k'}}{(1+aT)^2}~dT = 
\frac{_2F_1(1-k', -k'; 2-k';z)}{a^{1+k'}(k'-1)(1+aT)^{1-k'}}~,
\end{equation}
where $z = (1+aT)^{-1}$.
\begin{figure*}
\centering
\includegraphics[width=14.9cm, height=11.5cm]{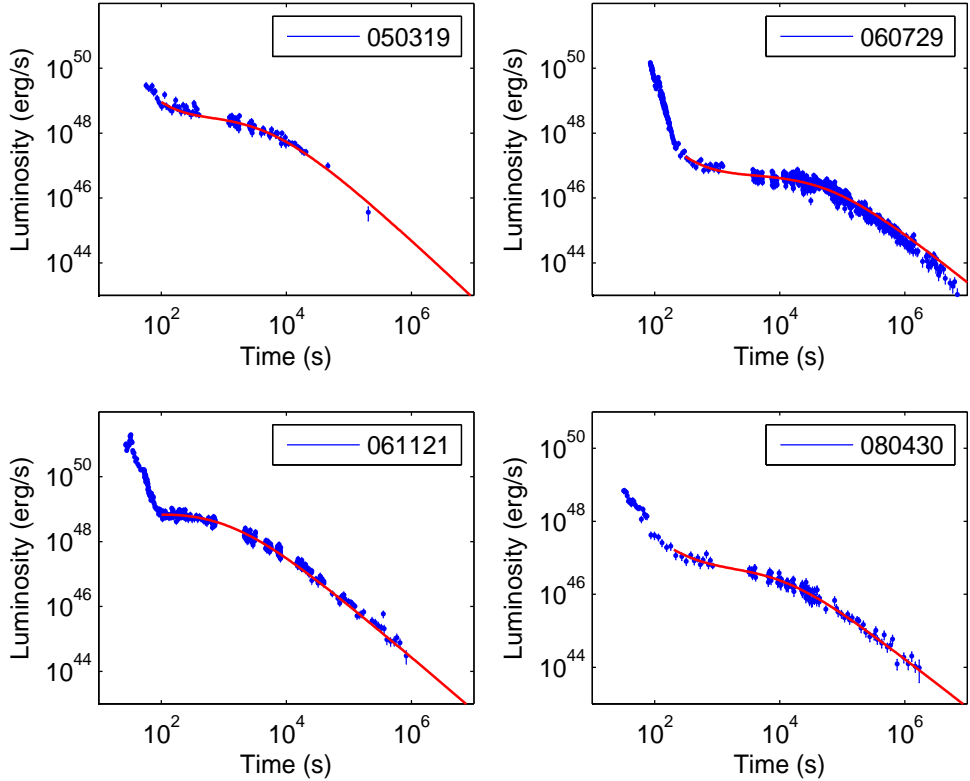}
\caption{Least square fit to the X-ray lightcurves of four selected GRB 
afterglows observed by Swift/XRT obtained through our eqs. 
(\ref{formally-solved}) and (\ref{hypergeom}). The blue points and 
red lines represent the measured (isotropic) luminosity (in the 0.3-100 keV
energy range) and best fits, respectively. The starting time T$_0$ was 
determined for each afterglow based on the end of the spectral transition from 
the previous steep-decay phase, as discussed in the text. Best-fit parameter 
values for each individual lightcurve are reported in Tab. (\ref{tab1}).}
\label{fig2}
\end{figure*}
%
Inserting this expression in the above eq. (\ref{formally-solved}), we 
obtain the complete functional form of $E(T)$
%
%
and can accordingly re-express the energy loss term of eq. (\ref{energytwo}) as 
$L(T) = k^{'}E(T)/T$, which represents the total (bolometric) luminosity of the 
blastwave. The resulting function thus provides the total bolometric luminosity 
as a function of (observer's) time $T$. In what follows we use it as an 
approximation to the X-ray lightcurve in order to compare it with the X-ray 
data. This assumes that the observed X-ray luminosity matches the bolometric 
luminosity, a reasonable approximation 
as long as the X-ray flux is the dominant emission component, 
which is largely verified in the early stages of the afterglow.
\section{Discussion}
\label{discussion}
In order to illustrate the salient properties of our solution, we plot in 
Fig. \ref{fig1} five lightcurves calculated through 
eqs. (\ref{formally-solved}) and (\ref{hypergeom}). These correspond to a 
choice of typical values for the shock initial energy, spin period and (dipole) 
magnetic field of the newly formed magnetar. For all curves a value $k'=0.4$ 
for the radiative efficiency was assumed. 

The curves in Fig. (\ref{fig1}) highlight some general features of the 
solution, which are 
not related to the nature of our approximation. 
In general, the model predictions depend mainly on
three key parameters: the NS initial spin energy, its (dipole) external 
magnetic field and the initial energy in the blast wave. The observed X-ray
lightcurves of GRB afterglows display a range of shapes, normalizations and 
durations which our model can account for in a very natural way. 
On average, the shallow phase is not flat but characterized by a
negative slope, smoothly steepening over time as energy injection decreases and 
radiative losses become more important. In fact, the NS spindown luminosity 
itself, \textit{i.e.} the energy injection term, has an effective power-law 
slope $\alpha_{\mbox{\tiny{sd}}}$ (to distinguish it from $\alpha$ of the observed 
afterglow lightcurve), which is expressed as
$\alpha_{\mbox{\tiny{sd}}} (t)=-2 (t/t_2)/(1+t/t_2)$. This equals zero at $t=0$ 
and gradually steepens with time, reaching the value
$\alpha_{\mbox{\tiny{sd}}} = 1$ at $t=t_2$ and, eventually, $\alpha_{\mbox{\tiny{sd}}} 
\rightarrow 2$ for $t \rightarrow \infty$. This allows a simple understanding
of the lightcurve shapes and their dependence on the parameters.

Given the form of the radiative loss term, $\sim E/T$, the condition $dE/dT > 
0$ is necessary and sufficient for the shallow decay phase to occur, namely
a part of the lightcurve where the temporal index of the power-law decay is
smaller than 1. Therefore, the onset of the shallow decay phase can be defined 
by the requirement $L_{\mbox{\tiny{inj}}}(T) > kE(T)/T$. This leads to a bound on 
the time at which the shallow phase starts, which can be expressed in terms of 
the ratio $x =T/T_2$ as
\begin{equation}
\label{bound}
\frac{x}{(1+x)^2} > k\frac{E(T)}{E_{s,i}}~.
\end{equation}
In general, at the initial stage of afterglow emission, the large value of E 
and small value of $T$ make it likely that radiative losses dominate over the 
injection term. Starting from the initial (kinetic) energy of the explosion, 
$E_0$, the afterglow luminosity will then decay as a power-law with index 
$(1+k)$. In this situation, one has $E(T) = E_0 (T_0/T)^k$ that, inserted in 
eq. (\ref{bound}) above, gives
\begin{equation}
\label{bound2}
\frac{x^{1+k}}{(1+x)^2} > k \frac{E_0}{E_{s,i}}\left(\frac{T_0}{T_2}\right)^k~.
\end{equation}
Putting $T=T_0 (\ll T_2)$ in this equation, and then defining $x_0 = T_0/T_2$, 
one sees that the shallow phase can start right at time $T_0$ if $x_0/(1+x_0)^2 
> kE_0/E_{s,i}$. However, since the left hand side in this inequality is $\ll 1$ 
by definition, the condition could be verified only for a very small ratio 
$E_0/E_{s,i}$. 
Although this might happen in some cases, one does not expect this to be 
a general occurrence, so that an initial power-law decay of the lightcurve, 
with index $\alpha = 1+k >1$ is to be expected in general. After 
some time, however, condition (\ref{bound2}) will be met so that energy 
injection will 
overcome radiative losses, the total energy within the shell will start 
increasing over time and the lightcurve will flatten accordingly. 

This shallow decay phase will clearly last as long as energy injection is 
sufficiently strong to balance radiative losses, a condition that is bound to
fail somewhere after time $T_2$. Indeed, we know that during the shallow phase 
$E(T)$ increases over time $\propto T^{\beta}$, with $\beta <1$ and, 
as long as $T < T_2$,   
$x$ on the left-hand side of eq. (\ref{bound}) grows $\sim T$. Hence, if
that inequality was satisfied at some early time, it will continue to hold at 
least up to $\sim T_2$. After time $T_2$, on the other hand, both the left-hand 
side and right-hand side of eq. (\ref{bound}) start decreasing with a 
steepening dependence on time. However, while the latter term has an asymptotic 
decay $\propto T^{-k}$ dictated by the form of radiative losses, the 
former term has a steeper asymptotic decay, $\propto T^{-1}$, dictated 
by the form of the injection term. Therefore, the inequality expressed in eq. 
(\ref{bound}) is 
bound to reverse sign at some time $T > T_2$, which implies $dE/dT <0$ at that
time and the afterglow luminosity will eventually decay again $\propto 
T^{-(1+k)}$.

A natural anticorrelation between the duration of the shallow phase and its 
luminosity, \textit{i.e.} the afterglow luminosity, is also apparent from our 
model lightcurves, again matching one of the salient properties of observed 
X-ray afterglows (Sato et al. 2007, Dainotti et al. 2008). In the 
light of the above discussion it is straightforward to see that this 
anticorrelation reflects an intrinsic property of the energy injection model  
assumed, \textit{i.e.} of magnetic dipole spindown. In fact, NSs with a larger 
spindown luminosity will, on average, have a shorter spindown timescale. This 
can be checked from the last step in eq. (\ref{Lsd}), in the limit $t<t_2$ 
appropriate for the shallow decay phase, showing that the NS initial spindown 
luminosity is $L_i \propto B^2 \omega^4_i$, while $t^{-1}_2 \propto B^2 
\omega^2_i$. 
 
This point also leads to the expectation that the most luminous among all 
GRBs/afterglows might even lack a shallow phase altogether. This expectation 
can also be seen directly from eq. (\ref{bound2}). The left-hand side of that 
equation has indeed a maximum at $x=(1+k)/(1-k)$ so, if the right-hand side was 
always greater than that maximum, the condition $dE/dT >0$ could never be met.
As an example, choosing $k=0.5$ gives a maximum $\simeq 0.325$ at $x=3$. 
Therefore, no shallow decay phase would occur if $E_0/E_{s,i} > 
0.65(T_2/T_0)^{1/2}$. From this example we see that, for a given initial spin 
energy, shorter values of $T_2$ (implying stronger magnetic fields) can more 
easily lead to a lack of the shallow decay phase. On the other hand, for a 
given $T_2$, smaller values of the initial spin energy (again implying stronger 
values of the magnetic field) favour the lack of a shallow phase.
 

Finally we note that, even starting
with largely different initial luminosities, theoretical lightcurves 
converge to a narrow distribution in luminosity at late times, nicely 
reproducing a property of observed lightcurves (cfr. Nousek et al. 2006, their 
Fig. 2). Again, this can be easily understood in terms of the above discussion.
All lightcurves evolve as $T^{-(1+k)}$ at late times, \textit{i.e.} for $t >> 
t_2$ when energy injection has long ceased, so that they are all parallel for 
a fixed value of $k'$.
Second, the general anticorrelation between luminosity and duration of 
energy injection implies that lower luminosity plateau's will, in general,
last longer than higher luminosity ones. 
Therefore the power-law decay starts earlier in more luminous afterglows, while 
less luminous ones are still (nearly) flat, which naturally causes all late
power-law segments to look as if they had very similar normalizations. 
Basically, this will reflect the overall energy budget of the blastwave, 
including both the initial kinetic energy and the energy injected in the later 
phase. Note that if the same total energy is injected over a longer time 
- hence, at a lower luminosity - it can produce a late power-law decay with a 
slightly higher normalization, as clear from the second and third curves in
Fig. \ref{fig1}. 
\subsection{Comparison with Observations and Results}
%
In order to further assess the goodness of our treatment, we show in Fig. 
(\ref{fig2}) fits to four selected X-ray afterglows observed by Swift, based on 
our solution (see eq. \ref{formally-solved}). 
The afterglows were chosen for illustration purposes among GRBs with known 
redshifts, good statistics in
the XRT lightcurves, clear evidence for a shallow phase, sufficiently long 
monitoring ($ > 10^5$ s after the trigger) and absence of bright flares or 
re-brightenings superposed on their lightcurves. 

To approximate bolometric luminosity, we obtained rest frame 0.3-100 keV light 
curves (Fig. 2) from the observed 0.3-10 keV counts rate (taken from Swift XRT 
lightcurve repository, see Evans et al. 2006, 2009), assuming an absorbed power 
law spectral model over the shallow phase, including also the subsequent normal 
decay if necessary. 
The resulting (isotropic) luminosities were calculated
by multiplying the fluxes by $4 \pi D_L^2$, with $D_L$ the luminosity distance 
calculated by assuming a standard $\Lambda$CDM cosmology with 
$\Omega_m=0.27$, $\Omega_{\Lambda}=0.73$ and $H_0=0.73$ km s$^{-1}$ Mpc$^{-1}$.
Finally, observer's times were corrected for the redshift of the source. 
In accordance with the idea that the initial steep decay phase results from the
prompt event, we only fitted data points from the onset of the shallow decay 
phase, as determined by the time
at which the spectral transition of the X-ray emission takes place (our
 T$_0$).

We note that, even though only data points for $T >T_0 \sim$ 100 s were 
fitted, the initial spin energy of our model refers to the time at which 
magnetic dipole braking sets in, which, in general, is earlier than T$_0$. 
Moreover, the magnetic field is assumed to remain constant. We stress that
our derived values of the radiative efficiency $k'$ should be taken with some
caution. As discussed in the previous section, the degeneracy between the 
different parameters that determine $k'$ can only be solved through detailed 
hydrodynamical models and joint multiwavelength fits to observations. Here we
note that best-fit values of $k'$ correspond to values of $\epsilon_e$ in a 
relatively wide, and fully acceptable, range between 0.1 and 0.8, depending on 
the exact relation between $t$ and $T$.
The results of our fits show that our model is in agreement with the data in 
our sample. NS parameters resulting from our least-squares fits are reported in 
Tab. (\ref{tab1}). We stress that the initial spin periods, $\sim 1-3$ ms, 
and the magnetic fields, several times 10$^{14}$ G up to slightly larger than
10$^{15}$ G, perfectly match the parameter range expected for newly born, 
millisecond spinning magnetars, if their seed magnetic fields were amplified
to ultrastrong values by a strong $\alpha-\Omega$ dynamo in the first seconds 
after formation (Duncan \& Thompson 1992). In addition, we note that the 
derived magnetic field values also perfectly match the values of the dipole 
fields measured in galactic X-ray sources thought to host magnetars, namely the 
Anomalous X-ray Pulsars and Soft-Gamma Ray Repeaters (see Woods \& Thompson 
2006, Mereghetti 2008). Finally, the required initial spin periods imply 
spin energies in the range ($3\times 10^{51} - 3 \times 10^{52}$) erg, for 
a NS moment of inertia I $\simeq 1.4 \times 10^{45}$ g cm$^2$ (cfr. 
Lattimer \& Prakash 2001).
Note that this is a very conservative estimate of the total spin energy of the 
NS at formation, since we are only measuring the spin energy of the NS as 
magnetic dipole losses set in as the dominant spindown mechanism.
Additional mechanisms have been proposed, which might be initially more 
efficient in extracting the NS spin energy and angular momentum, as long as  
it spins at $\simeq 1$ ms (cfr. Bucciantini et al. 2006, 2008, Dall'Osso, 
Stella \& Shore 2009, Metzger 2010). 
\begin{table}
\begin{center}
\begin{tiny}
\caption{Best-Fit Parameters for the four selected GRBs. The time T$_0$ is the 
time starting from which we fit observed lightcurves (see text for more 
details) and the energy E$_0$ respresents the total energy, in units of 10$^{50}$
erg, within the blastwave at time T$_0$. The dipole magnetic field of the NS is 
expressed in units of $10^{14}$. Reported errors are at 1$\sigma$.}\label{tab1}
\begin{tabular}{lccccc}
\hline
\hline
GRB & P$_i$ [ms] & B$_{14}$ [G] & k & E$_{0,50}$ [erg] &  $T_0$ [s] \\ 
\hline
~\\
050319 &  1.05 $\pm$ 0.03 & 5.5 $\pm$ 0.4 & 0.80 $\pm$ 0.25  & 10.0 $\pm$ 2.1 & 
100  \\
060729 & 2.12 $\pm$ 0.08 & 3.2 $\pm$ 0.2 & 0.58  $\pm$ 0.11 & 1.3 $\pm$ 0.2  & 
300 \\
061121 &   1.18 $\pm$  0.06 & 12.2 $\pm$ 0.7 & 0.66  $\pm$ 0.21 & 10.4 $\pm$ 
3.0  & 100 \\
080430 & 3.7 $\pm$ 0.6 & 11.7 $\pm$ 1.1 & 0.33 $\pm$ 0.09 & 1.2 $\pm$ 0.3 & 
200 \\
\hline
\hline
\end{tabular}
\label{tab1}
\end{tiny}
\end{center}
\end{table}
%
We finally note that in at least one case (GRB060729), the X-ray luminosity at 
late times is somewhat lower than the model prediction.
This can be caused by effects that have not been considered in our 
simple model, of which we breifly mention two. First, as we discussed in $\S$ 
\ref{model}, forcing the model to fit the whole lightcurve with a fixed value 
of dln$t$/dln$T$ can lead to an overestimate (by a factor of 2 at most) of the 
late time emission for a given shallow phase luminosity. 
This is just of the correct magnitude to explain the mismatch in the case of 
GRB060729. \\
Secondly, fitting 
the X-ray lightcurves with the bolometric luminosity can lead to an 
overestimate, particularly at late times. In fact the 
X-ray lightcurves probably are not representative of the bolometric 
luminosity at late times, when the contribution of other, lower frequency 
bands to the total emission becomes non-negligible.
\subsection{Further developments}
\label{further}
We briefly mention here some of the points that deserve futher comment and 
that we identify as the most important steps to improve the model beyond the
present treatment. 

As found by several authors (Panaitescu et al. 2006b, Liang et al. 
2007), the optical lightcurves of several GRB afterglows show a chromatic 
behaviour when compared to X-ray lightcurves, namely they do not show any break 
at the shallow-to-normal break observed in X-rays. According to Liang et al. 
(2007), about a half of GRB afterglows with simultaneous X-ray and optical 
observations show this chromatic behaviour. This is customarily viewed as a 
major problem for most models invoked to explain the shallow decay phase, at 
least for half of the afterglows.
As far as the model introduced here is concerned a self-consistent 
solution of $\Gamma(t)$ is required for a detailed calculation of the expected 
synchrotron emission at different frequencies (cfr. Zhang \& Meszaros 2004, 
Sari 2006). This will permit to derive 
multiband lightcurves up to late times which can be compared with 
multiwavelenght observations of GRB afterglows. However we stress here
that a 
different behavior of the
optical and X-ray lightcurves is to be expected, in general, in the 
synchrotron emission from the external shock, if a spectral break frequency 
were located between the X-ray and optical bands (as already noted by 
Panaitescu et al. 2006b). This is a natural property of a flow in the slow 
cooling regime, 
if the cooling frequency lies above the optical band but below the X-ray band 
during (most of) the energy injection phase. In this case, while electrons 
contributing to the X-ray emission are re-radiating instantaneously all the 
energy that is transferred to them, electrons emitting in the optical are 
radiating only a tiny fraction of it. Therefore, the optical emission can
evolve indipendently of the injection term for a while, until the emitting 
electrons nearly exhaust the energy that they accumulated. We will present 
a quantitative treatment of this issue in a future paper.

The assumed geometry of the expanding ejecta also deserves some discussion.
Although the subject is still open to debate, there exists by now 
considerable evidence that GRB fireballs are beamed, likely with opening angles 
of several degrees (Frail et al. 2001, Ghirlanda et al. 2004, Nava et al. 2007, 
Liang et al. 2008, Racusin et al. 2009, Cenko et a. 2009). 
It is important to realize that the
energetic requirements of the central engine as derived from our fits are not 
sensitive to the degree of beaming of the ejecta. This is true as long as the 
NS emits its spindown luminosity in a nearly isotropic way, as envisaged in 
models of NS magnetospheres loosing spin energy and angular momentum through 
the open field lines (Contopoulos, Kazanas \& Fendt 1999, Gruzinov 2005, 
Spitkovsky 2006).
In this case, only a fraction $\sim \theta^2_j$ of the rotational energy 
losses will be transferred to the ejecta and contribute to L$_{\mbox{\tiny{inj}}}$ 
in eq. (\ref{energyone}), where $\theta_j$ is the opening angle of the beam. On 
the other hand, the measured luminosity of the afterglow would have to be 
decreased by the same factor, thus leaving the inferred (isotropic) spindown 
luminosity of the NS unaffected. The characteristic timescale $t_2$ of energy 
injection is also unaffected by beaming, implying that our derived values of 
the dipole magnetic field are also independent of beaming in this case. 

On the contrary, if the millisecond spinning NS were to emit its spindown
luminosity in a beam of comparable size to the beam of the fireball, then
all its power would be transferred to the ejecta. The required spin energy 
of the NS ($\propto \omega^2_i$) would have to be reduced accordingly by 
a factor $\sim \theta^2_j$. In this case, since the timescale $t_2 \propto B^2 
\omega^2_i$ is always independent of beaming, our derived values of the 
magnetic field would have to be increased by a factor $\sim \theta^{-1}_j$.

\section{Conclusions}
\label{conclusions}
In the framework of prolonged energy injection models for GRB afterglows 
observed by Swift, we have considered the possibility that newly born 
magnetars - strongly magnetized and millisecond spinning NSs - are formed in 
the events producing (long) GRBs. In the first hours after formation of the 
NSs, the high spindown luminosity caused by magnetic dipole radiation losses 
represents a natural mechanism for prolonged energy injection in the external 
shock. To assess the viability of this scenario
%
we considered the energy balance of a blastwave subject to injection of 
energy by a NS spinning down through magnetic dipole radiation, along with 
radiative losses ($\propto E/t$). We found an approximate expression for the 
(isotropic) bolometric luminosity of the blastwave as a function of 
time that is in substantial agreement with general properties of 
the shallow-decay and normal-decay phases of X-ray GRB afterglows observed by 
Swift. 

Moreover, we have shown that individual lightcurves can be very well fitted 
by using our derived expression for the bolometric luminosity of the 
continuously-powered blastwave. In particular, our best fits provide values for 
the initial spin period of the NS in the range 1-3 ms, which match well the 
values expected in magnetar formation scenarios. Best-fit 
values for the magnetic dipole field,  $10^{14}-10^{15}$ G, are also in the 
range expected for such objects at formation and in agreement with the dipole 
fields estimated for Anomalous X-ray Pulsars and Soft Gamma-ray Repeaters, the
candidate magnetars in our Galaxy.

\begin{acknowledgements}
This work made use of data supplied by the UK Swift Science Data Centre at the 
University of Leicester.\\
SD acknowledges support from a VESF Fellowship. SD thanks Prof. S. N. Shore 
for insightful discussions and helpful comments. 
\end{acknowledgements}

\clearpage

\clearpage

\clearpage


\clearpage



\clearpage


\begin{thebibliography}{}

\bibitem[Barkov \& Komissarov(2009)]{2009MNRAS.tmp.1713B} Barkov, M.~V., \& 
Komissarov, S.~S.\ 2009, \mnras, 1713 
\bibitem[Bianco \& Ruffini(2004)]{2004ApJ...605L...1B} Bianco, C.~L., \& 
Ruffini, R.\ 2004, \apjl, 605, L1 
\bibitem[Blackman \& Yi(1998)]{1998ApJ...498L..31B} Blackman, E.~G., \& Yi, I.\ 
1998, \apjl, 498, L31 
\bibitem[Blandford \& McKee(1976)]{1976PhFl...19.1130B} Blandford, R.~D., \& 
McKee, C.~F.\ 1976, Physics of Fluids, 19, 1130 
\bibitem[Bucciantini et al.(2006)]{2006MNRAS.368.1717B} Bucciantini, N., 
Thompson, T.~A., Arons, J., Quataert, E., \& Del Zanna, L.\ 2006, \mnras, 368, 
1717 
\bibitem[Bucciantini et al.(2008)]{2008MNRAS.383L..25B} Bucciantini, N., 
Quataert, E., Arons, J., Metzger, B.~D., 
\& Thompson, T.~A.\ 2008, \mnras, 383, L25 
\bibitem[Bucciantini et al.(2009)]{2009MNRAS.396.2038B} Bucciantini, N., 
Quataert, E., Metzger, B.~D., Thompson, T.~A., Arons, J., 
\& Del Zanna, L.\ 2009, \mnras, 396, 2038 
\bibitem[Cannizzo \& Gehrels(2009)]{2009ApJ...700.1047C} Cannizzo, J.~K., \& 
Gehrels, N.\ 2009, \apj, 700, 1047
\bibitem[Cenko et al. 2009]{2009arxiv0905.0690}S. B. Cenko, D. A. Frail, 
F. A. Harrison, S. R. Kulkarni, E. Nakar, P. Chandra, N. R. Butler, D. B. Fox, 
A. Gal-Yam, M. M. Kasliwal, J. Kelemen, D.-S. Moon, P. A. Price, A. Rau, 
A. M. Soderberg, H. I. Teplitz, M. W. Werner, D. C.-J. Bock, J. S. Bloom, 
D. A. Starr, A. V. Filippenko, R. A. Chevalier, N. Gehrels, J. N. Nousek, 
T. Piran .\ 2009, arXiv:0905.0690
\bibitem[Cohen et al.(1998)]{1998ApJ...509..717C} Cohen, E., Piran, T., 
\& Sari, R.\ 1998, \apj, 509, 717 
\bibitem[Contopoulos et al.(1999)]{1999ApJ...511..351C} Contopoulos, I., 
Kazanas, D., \& Fendt, C.\ 1999, \apj, 511, 351 
\bibitem[Corsi \& M{\'e}sz{\'a}ros(2009)]{2009ApJ...702.1171C} Corsi, A., \& 
M{\'e}sz{\'a}ros, P.\ 2009, \apj, 702, 1171 
\bibitem[Cutler(2002)]{2002PhRvD..66h4025C} Cutler, C.\ 2002, \prd, 66, 
084025 
\bibitem[Dai \& Lu(1998)]{1998A&A...333L..87D} Dai, Z.~G., \& Lu, T.\ 1998, 
\aap, 333, L87
\bibitem[Dainotti et al.(2008)]{2008MNRAS.391L..79D} Dainotti, M.~G., 
Cardone, V.~F., \& Capozziello, S.\ 2008, \mnras, 391, L79  
\bibitem[Dall'Osso \& Stella(2007)]{2007Ap&SS.308..119D} Dall'Osso, S., \& 
Stella, L.\ 2007, \apss, 308, 119 
\bibitem[Dall'Osso et al.(2009)]{2009MNRAS.398.1869D} Dall'Osso, S., Shore, 
S.~N., \& Stella, L.\ 2009, \mnras, 398, 1869 
\bibitem[de Pasquale et al.(2007)]{2007MNRAS.377.1638D} de Pasquale, M., et 
al.\ 2007, \mnras, 377, 1638 
\bibitem[Duncan \& Thompson(1992)]{1992ApJ...392L...9D} Duncan, R.~C., \& 
Thompson, C.\ 1992, \apjl, 392, L9
\bibitem[Evans et al.(2007)]{2007A&A...469..379E} Evans, P.~A., et al.\ 2007, 
\aap, 469, 379 
\bibitem[Evans et al.(2009)]{2009MNRAS.397.1177E} Evans, P.~A., et al.\ 
2009, \mnras, 397, 1177  
\bibitem[Fan \& Xu(2006)]{2006MNRAS.372L..19F} Fan, Y.-Z., \& Xu, D.\ 2006, 
\mnras, 372, L19 
\bibitem[Frail et al.(2001)]{2001ApJ...562L..55F} Frail, D.~A., et al.\ 
2001, \apjl, 562, L55 
\bibitem[Gehrels et al.(2009)]{2009ARA&A..47..567G} Gehrels, N., Ramirez-Ruiz, 
E., \& Fox, D.~B.\ 2009, \araa, 47, 567 
\bibitem[Ghirlanda et al.(2004)]{2004ApJ...613L..13G} Ghirlanda, G., 
Ghisellini, G., Lazzati, D., \& Firmani, C.\ 2004, \apjl, 613, L13 
\bibitem[Grupe et al.(2007)]{2007ApJ...662..443G} Grupe, D., et al.\ 
2007, \apj, 662, 443
\bibitem[Gruzinov(2005)]{2005PhRvL..94b1101G} Gruzinov, A.\ 2005, Physical 
Review Letters, 94, 021101 
\bibitem[Klu{\'z}niak \& Ruderman(1998)]{1998ApJ...505L.113K} Klu{\'z}niak, W., 
\& Ruderman, M.\ 1998, \apjl, 505, L113
\bibitem[Liang et al.(2007)]{2007ApJ...670..565L} Liang, E.-W., Zhang, 
B.-B., \& Zhang, B.\ 2007, \apj, 670, 565 
\bibitem[Liang et al.(2008)]{2008ApJ...675..528L} Liang, E.-W., Racusin, 
J.~L., Zhang, B., Zhang, B.-B., \& Burrows, D.~N.\ 2008, \apj, 675, 528 
\bibitem[Lyons et al.(2009)]{2009arXiv0908.3798L} Lyons, N., O'Brien, 
P.~T., Zhang, B., Willingale, R., Troja, E., \& Starling, R.~L.~C.\ 2009, 
arXiv:0908.3798 
\bibitem[MacFadyen et al.(2001)]{2001ApJ...550..410M} MacFadyen, A.~I., 
Woosley, S.~E., \& Heger, A.\ 2001, \apj, 550, 410 
\bibitem[Mereghetti(2008)]{2008A&ARv..15..225M} Mereghetti, S.\ 2008, \aapr, 
15, 225 
\bibitem[Meszaros et al.(1999)]{1999NewA....4..303M} Meszaros, P., Rees, 
M.~J., \& Wijers, R.~A.~M.~J.\ 1999, New Astronomy, 4, 303 
\bibitem[Metzger(2010)]{2010arXiv1001.5046M} Metzger, B.~D.\ 2010, 
arXiv:1001.5046 
\bibitem[Narayan et al.(1992)]{1992ApJ...395L..83N} Narayan, R., Paczynski, 
B., \& Piran, T.\ 1992, \apjl, 395, L83 
\bibitem[Nava et al.(2007)]{2007MNRAS.377.1464N} Nava, L., Ghisellini, G., 
Ghirlanda, G., Cabrera, J.~I., Firmani, C., 
\& Avila-Reese, V.\ 2007, \mnras, 377, 1464 
\bibitem[Palomba(2001)]{2001A&A...367..525P} Palomba, C.\ 2001, \aap, 
367, 525 
\bibitem[Paczynski \& Xu(1994)]{1994ApJ...427..708P} Paczynski, B., \& Xu, G.\ 
1994, \apj, 427, 708 
\bibitem[Panaitescu et al.(2006)]{2006MNRAS.366.1357P} Panaitescu, A., 
M{\'e}sz{\'a}ros, P., Gehrels, N., Burrows, D., 
\& Nousek, J.\ 2006a, \mnras, 366, 1357 
\bibitem[Panaitescu et al.(2006)]{2006MNRAS.369.2059P} Panaitescu, A., 
M{\'e}sz{\'a}ros, P., Burrows, D., Nousek, J., Gehrels, N., O'Brien, P., 
\& Willingale, R.\ 2006b, \mnras, 369, 2059 
\bibitem[Piran(1999)]{1999PhR...314..575P} Piran, T.\ 1999, \physrep, 314, 
575 
\bibitem[Racusin et al.(2009)]{2009ApJ...698...43R} Racusin, J.~L., et al.\ 
2009, \apj, 698, 43 
\bibitem[Ramirez-Ruiz(2004)]{2004MNRAS.349L..38R} Ramirez-Ruiz, E.\ 2004, 
\mnras, 349, L38 
\bibitem[Rees \& Meszaros(1994)]{1994ApJ...430L..93R} Rees, M.~J., \& Meszaros, 
P.\ 1994, \apjl, 430, L93 
\bibitem[Rees \& Meszaros(1998)]{1998ApJ...496L...1R} Rees, M.~J., \& 
Meszaros, P.\ 1998, \apjl, 496, L1 
\bibitem[Sari \& Piran(1997)]{1997MNRAS.287..110S} Sari, R., \& Piran, T.\ 
1997, \mnras, 287, 110 
\bibitem[Sari et al.(1998)]{1998ApJ...497L..17S} Sari, R., Piran, T., 
\& Narayan, R.\ 1998, \apjl, 497, L17 
\bibitem[Sari \& M{\'e}sz{\'a}ros(2000)]{2000ApJ...535L..33S} Sari, R., \& 
M{\'e}sz{\'a}ros, P.\ 2000, \apjl, 535, L33 
\bibitem[Sari(2006)]{2006AIPC..856...33S} Sari, R.\ 2006, Relativistic 
Jets: The Common Physics of AGN, Microquasars, and Gamma-Ray Bursts, 856, 
33 
\bibitem[Sato et al.(2007)]{2007arXiv0711.0903S} Sato, R., Ioka, K., Toma, 
K., Nakamura, T., Kataoka, J., Kawai, N., \& Takahashi, T.\ 2007, 
arXiv:0711.0903 
\bibitem[Spitkovsky(2006)]{2006ApJ...648L..51S} Spitkovsky, A.\ 2006, 
\apjl, 648, L51 
\bibitem[Stella et al.(2005)]{2005ApJ...634L.165S} Stella, L., Dall'Osso, 
S., Israel, G.~L., \& Vecchio, A.\ 2005, \apjl, 634, L165 
\bibitem[Taub(1948)]{1948PhRv...74..328T} Taub, A.~H.\ 1948, Physical 
Review , 74, 328 
\bibitem[Tchekhovskoy et al.(2009)]{2009ApJ...699.1789T} Tchekhovskoy, A., 
McKinney, J.~C., \& Narayan, R.\ 2009, \apj, 699, 1789 
\bibitem[Thompson \& Duncan(1993)]{1993ApJ...408..194T} Thompson, C., \& 
Duncan, R.~C.\ 1993, \apj, 408, 194 
\bibitem[Thompson \& Duncan(1995)]{1995MNRAS.275..255T} Thompson, C., \& 
Duncan, R.~C.\ 1995, \mnras, 275, 255 
\bibitem[Thompson \& Duncan(1996)]{1996ApJ...473..322T} Thompson, C., \& 
Duncan, R.~C.\ 1996, \apj, 473, 322
\bibitem[Thompson \& Duncan(2001)]{2001ApJ...561..980T} Thompson, C., \& 
Duncan, R.~C.\ 2001, \apj, 561, 980 
\bibitem[Thompson et al.(2004)]{2004ApJ...611..380T} Thompson, T.~A., 
Chang, P., \& Quataert, E.\ 2004, \apj, 611, 380 
\bibitem[Troja et al.(2007)]{2007ApJ...665..599T} Troja, E., et al.\ 2007, 
\apj, 665, 599 
\bibitem[Usov(1992)]{1992Natur.357..472U} Usov, V.~V.\ 1992, \nat, 357, 472
\bibitem[Wheeler et al.(2000)]{2000ApJ...537..810W} Wheeler, J.~C., Yi, I., 
H{\"o}flich, P., \& Wang, L.\ 2000, \apj, 537, 810  
\bibitem[Woods \& Thompson(2006)]{2006csxs.book..547W} Woods, P.~M., \& 
Thompson, C.\ 2006, ``Soft gamma repeaters and anomalous X-ray pulsars: 
magnetar candidates'',  Compact stellar X-ray sources, 547, eds. 
Lewin, W.~H.~G.~\& van der Klis, M. 
\bibitem[Woosley(1993)]{1993ApJ...405..273W} Woosley, S.~E.\ 1993, \apj, 
405, 273 
\bibitem[Xu et al.(2009)]{2009arXiv:0909.5318} Ming Xu, Yong-Feng Huang, 
Tan Lu .\ 2009,  arXiv:0909.5318
\bibitem[Yu \& Huang(2007)]{2007ChJAA...7..669Y} Yu, Y., \& Huang, Y.-F.\ 2007, 
Chinese Journal of Astronomy and Astrophysics, 7, 669 
\bibitem[Zhang \& M{\'e}sz{\'a}ros(2001)]{2001ApJ...552L..35Z} Zhang, B., \& 
M{\'e}sz{\'a}ros, P.\ 2001, \apjl, 552, L35 
\bibitem[Zhang \& M{\'e}sz{\'a}ros(2004)]{2004IJMPA..19.2385Z} Zhang, B., \& 
M{\'e}sz{\'a}ros, P.\ 2004, International Journal of Modern Physics A, 19, 2385 

\bibitem[Zhang et al.(2006)]{2006ApJ...642..354Z} Zhang, B., Fan, Y.~Z., 
Dyks, J., Kobayashi, S., M{\'e}sz{\'a}ros, P., Burrows, D.~N., Nousek, 
J.~A., \& Gehrels, N.\ 2006, \apj, 642, 354 
\bibitem[Zhang(2007)]{2007ChJAA...7....1Z} Zhang, B.\ 2007, Chinese Journal 
of Astronomy and Astrophysics, 7, 1 




















\end{thebibliography}
\end{document}